\documentclass[floatfix,aps,amsmath,nofootinbib,onecolumn,10pt]{revtex4}

\usepackage{graphicx}
\usepackage{bm}
\usepackage{rotating}
\usepackage{array}

\def\({\left(}
\def\){\right)}
\def\[{\left[}
\def\]{\right]}

\def\e{\begin{equation}}
\def\q{\end{equation}}
\def\m{\begin{eqnarray}}
\def\n{\end{eqnarray}}

\begin{document}

\title{Constraints on Newton's Constant from Cosmological Observations}

\author{Ke Wang$^{1}$ \footnote{wangkey@lzu.edu.cn} and Lu Chen$^{2}$ \footnote{chenlu@mail.itp.ac.cn}$^,$ \footnote{Corresponding author}}
\affiliation{$^1$ Institute of Theoretical Physics \& Research Center of Gravitation,\\ Lanzhou University, Lanzhou 730000, China\\
$^2$ School of Physics and Electronics, \\Shandong Normal University, Jinan 250014, China\\
}

\date{\today}

\begin{abstract}

Newton's constant has observational effects on both the CMB power spectra and the light curves of SNIa.
We use Planck data, BAO data and the SNIa measurement to constrain the varying Newton's constant $G$ during the CMB epoch and the redshift ranges of PANTHEON samples, and find no evidence indicating that $G$ is varying with redshift.
By extending the $\Lambda$CDM model with one free parameter $G$, we get $G =(6.65635_{-0.18560}^{+0.18766} ) \times 10^{-11} \rm m^3kg^{-1}s^{-2}$ and $H_0=67.62^{+1.24}_{-1.25} $ km s$^{-1}$ Mpc$^{-1}$ at 68$\%$ C.L. from Planck$+$BAO$+$uncalibrated PANTHEON. 
The results show the value of $G$ is consistent with CODATA 2018, but the $H_0$ tension can't be solved in this way.

\end{abstract}

\pacs{???}

\maketitle


\section{Introduction} 
\label{sec:intro}


Newton's gravitational constant is treated as a constant both in Newton's gravitational theory and general relativity.
Over one hundred years after Newton proposed its definition, Henry Cavendish measured the value of $G=6.754\pm 0.041 \times 10~\rm N \cdot \rm m^2/\rm k \rm g^2$ with torsion scale experiment.
Since then, kinds of methods are used to determine Newton's constant more precisely.
In 2019, the Committee on Data for Science and Technology (CODATA) gives its recommended value of $G =6.67430 \times 10^{-11}~\rm m^3\rm kg^{-1}\rm s^{-2}$ (named CODATA 2018) and the standard uncertainty is  $1.5 \times 10^{-15}~\rm m^3kg^{-1}s^{-2}$, which means $2.2 \times 10^{-5}$ relative uncertainty. 
In the laboratory, cold atom interferometry is also used to detect Newton's constant~\cite{Rosi:2014kva}.
In cosmology, the cosmic microwave background (CMB)~\cite{Umilta:2015cta,Ballardini:2016cvy,Zahn:2002rr,Galli:2009pr,Bai:2015vca,Xue:2014kna}, big bang nucleosynthesis (BBN)~\cite{Gelmini:2020ekg,Copi:2003xd,Alvey:2019ctk}, type Ia supernovae (SNIa)~\cite{Wright:2017rsu,Zhao:2018gwk,Riazuelo:2001mg,Zhang:2017aqn,Dhawan:2017ywl} and gravitational waves~\cite{Zhao:2018gwk,Vijaykumar:2020nzc} can provide different measurements of Newton's constant at corresponding epochs of our universe.
Obviously, there is a problem whether Newton's constant is always a constant really or not.
Theoretically, it is acceptable to be both time- or space-dependent in some theories of modified gravity~\cite{Rossi:2019lgt}. For example, the scalar-tensor theories predict a time-dependent $G$.
The cosmological observation provides a method to study the Newton's constant varying with redshift.

Any change in Newton's constant have influence on the expansion history of our universe, especially at the redshift of recombination, which leaving a footprint on the CMB power spectra.
Combing the precise observation of Planck collaboration~\cite{Aghanim:2018eyx}, Newton's constant during the CMB epoch can be restricted.
SNIa measurement, as the standard candles, are usually used to study the accelerated expansion, too.
Newton's constant affects its peak luminosity through the Chandrasekhar mass by $M_{\rm Ch} \propto G^{-3/2}$ mostly.
The latest SINa data, PANTHEON samples~\cite{Scolnic:2017caz}, detected the light curves of 1048 SNIa covering the redshift range $0<z<2.3$ and provides a way to limit Newton's constant at low redshift.
Therefore, we constrain the varying Newton's constant with the CMB power spectra and the SNIa peak luminosity and probe its dynamics. 

Besides, the Hubble constant $H_0$ indicates the expansion of the universe directly.
Plank collaboration claimed $H_0= 67.4 \pm 0.5 $ km s$^{-1}$ Mpc$^{-1}$ after its final data release~\cite{Aghanim:2018eyx}.
However, the SH0ES project yielded the best estimate as $H_0= 74.03 \pm 1.42 $ km s$^{-1}$ Mpc$^{-1}$ (named R19), which is 4.4$\sigma$ different from Planck~\cite{Riess:2019cxk}.
$H_0$ tension may result from systematic errors of measurements.  Errors of both the SH0ES and Planck data are studied in recent years～\cite{Lattanzi:2016dzq,Huang:2018xle,Spergel:2013rxa}. However, other alternative data show a similar discrepancy with the CMB measurement.
Another possibility is that $H_0$ tension implies new physics beyond the $\Lambda$CDM model.
Some experts attempt to solve the tension by extending the base $\Lambda$CDM model simply, such as the dark energy (DE) equation of state $w$, the effective number of relativistic species $N_{\text{eff}}$, the total mass of neutrinos $\Sigma m_{\nu}$, and so on~\cite{DiValentino:2016hlg,Chen:2017ayg,Vagnozzi:2019ezj,Kreisch:2019yzn,Hart:2019dxi,Ballardini:2020iws}.
Moreover, modifying early universe physics and changing late-time cosmology influence Hubble constant significantly.
From this view, dynamical DE~\cite{DiValentino:2017zyq,Yang:2018qmz,Li:2018nlh,Zhao:2017cud,Qing-Guo:2016ykt,Zhang:2019cww}, early DE~\cite{Braglia:2020bym,Poulin:2018cxd,Smith:2019ihp,Karwal:2016vyq}, interacting DE~\cite{Kumar:2016zpg,Yang:2018euj,DiValentino:2017iww,Begue:2017lcw}, dark radiation~\cite{Mortsell:2018mfj,Ko:2016uft}, scalar fields~\cite{Agrawal:2019lmo,Lin:2019qug}, and many other components are considered to solve $H_0$ tension.
Unfortunately, it has not been well solved till now.
Owing to the effect of Newton's constant on the Hubble parameter $H(z)$, we expect a solution of $H_0$ tension by modifying $G$.
Rencetly, Ref.~\cite{Ballesteros:2020sik} has discussed the varying $G$ in the scalar-tensor theory of gravity, which influences the expansion history of our universe before recombination epoch.
They gave the result of $H_0=69.2_{-0.75}^{+0.62}$ km s$^{-1}$ Mpc$^{-1}$.
And Ref.~\cite{Braglia:2020iik} also finds a larger value for $H_0$ by an evolving gravitational constant.
Here, we simply set $G$ as a free parameter in the base $\Lambda$CDM model to enlarge the value of $H_0$.

This paper is organized as follows.
In section \ref{sec:cmb}, we rescale Newton's constant by introducing $\lambda_2$ and sketch out its influence on the CMB power spectra.
In section \ref{sec:sne}, the effect of Newton's constant on the peak luminosity of SNIa is presented.
We show our results in section \ref{sec:results}.
We turn to CAMB and the Markov Chain Monte Carlo (MCMC) package CosmoMC~\cite{Lewis:2002ah}.
The CMB data, BAO data and the SNIa measurement are used to constrain the varying Newton's constant in section \ref{subsec:result-1}.
Then the data combination of CMB, BAO and uncalibrated SNIa is applied to study the $H_0$ tension with a constant G in section \ref{subsec:result-2}.
 Finally, a brief summary and discussion are included in section \ref{sec:sum}.

\section{Effects of Newton's Gravitational Constant on Cosmological Observations}
\label{sec:G}
To weigh the effects of Newton's gravitational constant on some cosmological observations, we rescale $G_N =6.6738 \times 10^{-11}~\rm m^3kg^{-1}s^{-2}$ with several dimensionless parameter $\lambda_i$, $i=0,1,2$, then the new definition of Newton's constant is
\begin{eqnarray}G=
\begin{cases}
\lambda_0^2G_N,&~\text{for}~z<0.1;\\
\lambda_1^2G_N,&~\text{for}~0.1\leq z<2.3;\\
\lambda_2^2G_N,&~\text{for}~2.3\leq z.\\
\end{cases}
\end{eqnarray}
Here, $G$ indicates the effective values of Newton's constant for each bins actually.
Then the Friedmann equation is
\begin{eqnarray}\mathcal{H}^2=\left(\frac{\dot{a}}{a}\right)^2=
\begin{cases}
\dfrac{8\pi}{3}a^2 \lambda_0^2 G_N \rho,&~\text{for}~z<0.1;\\
\dfrac{8\pi}{3}a^2 \lambda_1^2 G_N \rho,&~\text{for}~0.1\leq z<2.3;\\
\dfrac{8\pi}{3}a^2 \lambda_2^2 G_N \rho,&~\text{for}~2.3\leq z,\\
\end{cases}
\end{eqnarray}
where $\mathcal{H}$ is the Hubble rate , $a$ is the scale factor, $\rho$ is the total energy density in the universe and overdot means the differentiation over the conformal time $\tau$. When we rescale $\tau$ as
\begin{eqnarray}
d\tau \to \lambda_i d\tau=\frac{\lambda_i dt}{a}=\frac{da}{a^2 \sqrt{8\pi/3 G_N\rho}},
\end{eqnarray}
the integrand of cosmic distances are independent of $\lambda_i$. Therefore, we cannot use the cosmic distances only, like the BAO measurements with Eisenstein's baryon drag epoch $z_d$~\cite{Eisenstein:1997ik}, to constrain Newton's gravitational constant, but we can turn to other non-gravity interactions to constrain $G$.

\subsection{Effects of Newton's Gravitational Constant on CMB during the Recombination}
\label{sec:cmb}
To the first order, the Boltzmann equations of the baryons and photons in the conformal Newtonian gauge reads
\begin{eqnarray}
\label{eq:boltz}
     \dot{\delta}_\gamma &=& -{4\over 3}\theta_\gamma
	+4\dot{\phi} \,,\nonumber\\
     \dot{\theta}_\gamma &=& \frac{1}{4} k^2 \delta_\gamma
	+ k^2 \psi
	+ a n_e \sigma_T (\theta_b-\theta_\gamma) \,,\nonumber\\
	\dot{\delta}_b &=& -\theta_b + 3\dot{\phi} \,, \nonumber\\
	\dot{\theta}_b &=& -{\dot{a}\over a}\theta_b
	+ c_s^2 k^2\delta_b + {4\bar\rho_\gamma \over 3\bar\rho_b}
	 an_e\sigma_T (\theta_\gamma-\theta_b) + k^2\psi\,,\nonumber\\
 \end{eqnarray}
where $\delta = \delta \rho /\bar{\rho}$ is the density fluctuation, $\theta$ is the velocity perturbation for a given mode $k$, $\phi$ and $\psi$ represent the scalar mode of metric perturbations, $\sigma_T$ is the cross-section of Thomson scattering, $n$ is the number density and $(c_s^2)^{-1}=3\left(1+\dfrac{3 \bar\rho_b}{4\bar\rho_{\gamma}}\right)$ is the sound speed of baryons.
The subscript $e$ represents electrons, $\gamma$ is photons and $b$ means baryons.
If the two Thomson scattering terms in Eq.~(\ref{eq:boltz}) are ignored, the replacement of $\tau$ by $\lambda_i \tau $ must accompanies a replacement of $k$ by $k/\lambda_i$ for keeping Eq.~(\ref{eq:boltz}) (or CMB observations) unchanged. 
Therefore, the transformation of $\tau \to \lambda_i\tau$ also cannot be observed through perturbations because the transformation of $k \to k/\lambda_i$ can be compensated by adjusting the scalar spectral index $n_s$ appropriately if large-scale structure clustering measurements are not considered.
Fortunately, there exists Coulomb interaction. So the only way that $\lambda_i$ influences the CMB anisotropy spectrum is affecting the number of free electrons $n_e$ during the recombination epoch, hence the ionization fraction $x_e= n_e/n_{\rm H}=x_p+x_{\rm HeII}$ during the same epoch.
Here, $n_{\rm H}$ is the total number density of H nuclei, $x_p$ is the ionization fraction of H and $x_{\text{HeII}}$ presents that of He.
According to Ref.~\cite{Seager:1999bc}, the modified evolution of $x_p$ and $x_{\rm HeII}$ can be obtained by solving the following ordinary differential equations (ODEs)
\begin{eqnarray}
\frac{d x_p}{dz}&=&\frac{f_1(x_e,x_p,n_{\rm H},T_{\rm M})}{H(\lambda_2,z)(1+z)},\\
\frac{d x_{\rm HeII}}{dz}&=&\frac{f_2(x_e,x_{\rm HeII},n_{\rm H},T_{\rm M})}{H(\lambda_2,z)(1+z)},\\
\frac{d T_{\rm M}}{dz}&=&\frac{f_3(x_e,T_{\rm M},T_{\rm R})}{H(\lambda_2,z)(1+z)}+\frac{2T_{\rm M}}{(1+z)},
\end{eqnarray}
where $T_{\rm M}$ (or $T_{\rm R}$) is the matter (or radiation) temperature, the specific expressions of $f_1$, $f_2$ and $f_3$ are given in Ref.~\cite{Seager:1999bc}.
From above ODEs, we can find that $x_e$ evolves slower for $\lambda_2>1$, hence a latter photo-decoupling time $z_*$ and baryon drag epoch $z_d$. Therefore, we can use the data combination of CMB and BAO measurements to constrain $\lambda_2$.

\subsection{Effects of Newton's Gravitational Constant on the SNIa}
\label{sec:sne}
Since the effects of $\lambda_2$ on CMB is confined to $n_e$ during the recombination epoch, there is a possibility that the deviation of Newton's constant from $G_N$ is not equal to $\lambda_2$ at other different epochs. Therefore, it's necessary to introduce new parameters to quantify the potential deviation from $G_N$ after the recombination epoch, especially if the constraints on the deviation are not from CMB observations.

\begin{figure}[]
\begin{center}
\includegraphics[scale=0.45]{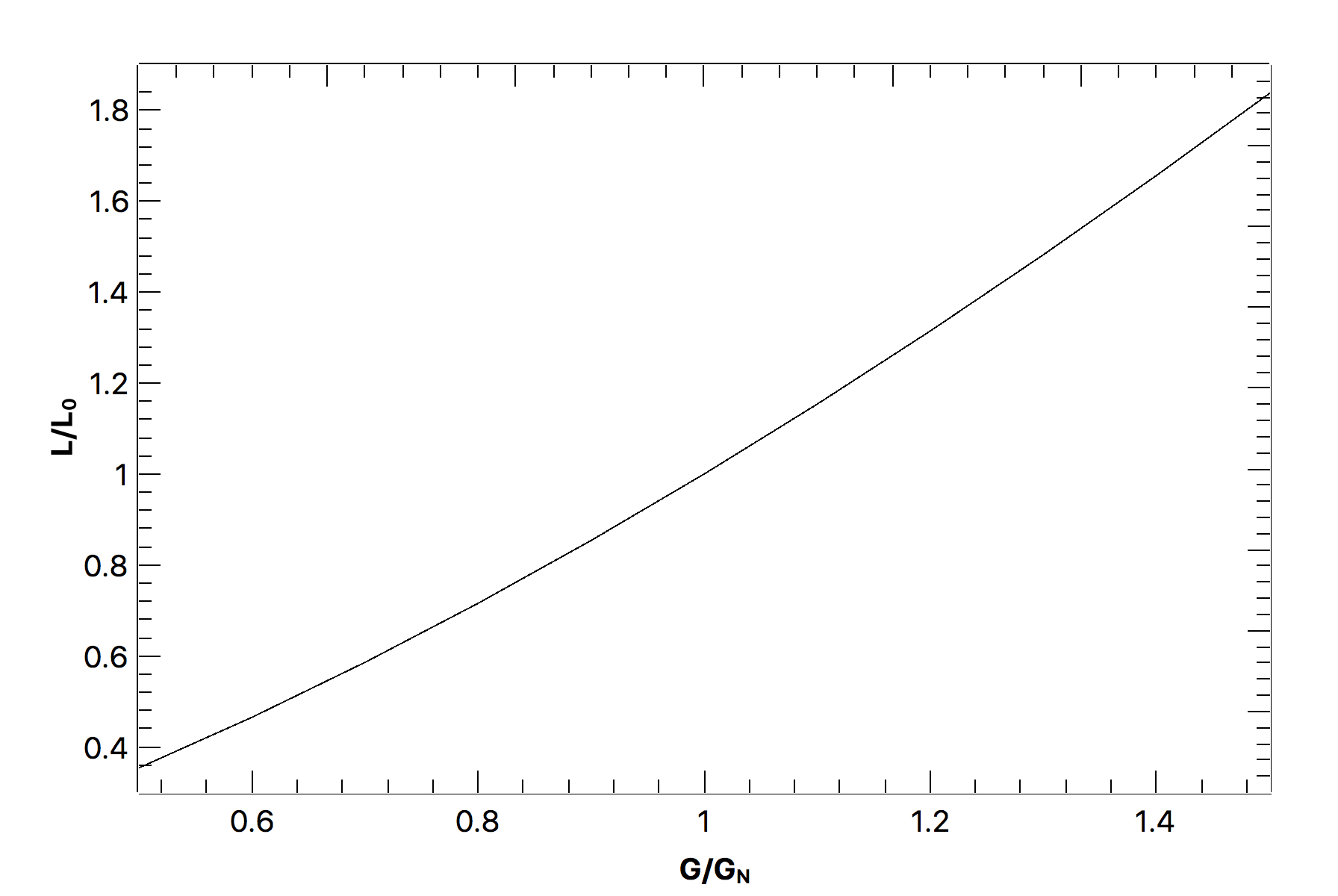}
\end{center}
\caption{$L/L_0$ as a function of $G/G_N$.}
\label{fig:LG}
\end{figure}

Newton's constant influences the light curve of SNIa via the Chandrasekhar mass $M_{\rm Ch} \propto G^{-3/2}$ mainly.
If the Newton's constant $G$ increases, the peak luminosity of light curve $L$ raises and its width drops~\cite{Wright:2017rsu}.
In this section, we introduce a new parameter $\lambda'_1= L/L_0$ to quantify the deviation of $L$ from $L_0$ resulting from the deviation of $G$ from $G_N$. 
Fig.~\ref{fig:LG} shows a sketch of $\lambda'_1=L/L_0$ as a function of $G/G_N$, from which we can derive $G/G_N$ from any given $\lambda'_1$.
The two parameters are almost linearly.
In other words, the derivation of $\lambda'_1$ from 1 is nearly equivalent to the difference between $G$ and $G_N$.
If we use the final redshifts, corrected magnitudes $\mu+M_B$ and the host galaxy mass $M_{\text{host}}$ of PANTHEON samples~\cite{Scolnic:2017caz} to constrain cosmological parameters, the combination of $\mu+M_B$ can be related to $\lambda'_1$ as
\begin{eqnarray}
\label{eq:mod}
\nonumber
\mu+M_B&=&5\log_{10}\big[\dfrac{d_L}{\rm Mpc}\big]+25+M_B^1-2.5\log_{10}\lambda'_1+\Delta_M\\
       &=&5\log_{10}\big[\dfrac{d_L}{\rm Mpc}\big]+25+M_{\odot}-2.5\log_{10} \big[\lambda'_1L_0/L_{\odot}\big] +\Delta_M
\end{eqnarray}
and $\Delta_M$ is related to $M_{\text{host}}$,
\begin{eqnarray}\Delta_M=
\begin{cases}
0,& \text{for}~M_\text{{host}}<10^{10} M_{\odot}, \\-0.08 \text{mag},& \text{for}~M_\text{{host}}\ge 10^{10} M_{\odot}.\\
\end{cases}
\end{eqnarray}
Due to the degeneracy between $\lambda'_1$ and $L_0$ (or $M_B^1$), we ignore the term of $-2.5\log_{10}\lambda'_1$ for $z<0.1$ and use samples at this redshift span to constrain $L_0$ (or $M_B^1$)\footnote{Since $L_0$ (or $M_B^1$) is defined with respect to $G_N$, it's convenient to set $\lambda_0=1$ to constrain $L_0$ (or $M_B^1$) directly.}. 
Then samples from $z>0.1$ will be used to constrain $\lambda'_1$. 

\section{Results}
\label{sec:results} 

\subsection{Varying G with Redshift}
\label{subsec:result-1}

\begin{figure}[]
\begin{center}
\includegraphics[scale=0.3]{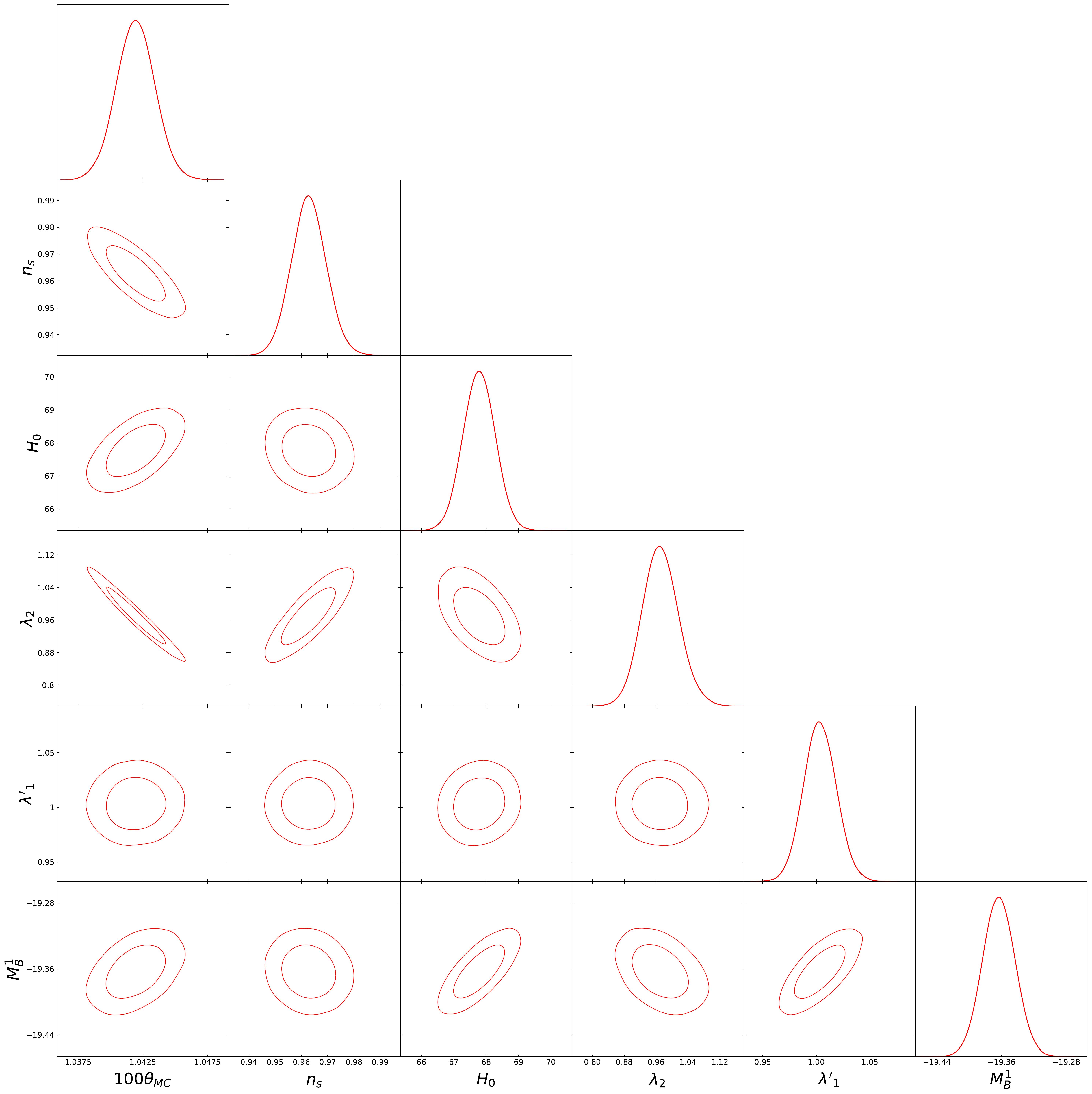}
\end{center}
\caption{The constraints on $\lambda'_1$, $\lambda_2$ and $M_B^1$ from the data combination of CMB, BAO and SNIa. Also we present the constraints on $n_s$ $H_0$ and $\theta_{\rm MC}$ which are affected most by the former three parameters.}
\label{fig:lambda12}
\end{figure}

Firstly, we consider an extension of $\Lambda$CDM model with another two free parameters $\lambda'_1$ (or $\lambda_1$) and $\lambda_2$ to probe the dynamics of $G$. 
Based on the previous discussion, $\lambda'_1$ and $\lambda_2$ are used to measure the varying Newton's constant $G$ during the period of SNIa measurement ($z\sim 0.1-2.3$) and the recombination epoch ($z\sim 1100$) respectively.
In summarize, the free parameters needed to be fitted are $\{ \Omega_b h^2,\Omega_c h^2, 100\theta_{\text{MC}}, \tau_{\text{re}}, \ln(10^{10}) A_s, n_s, \lambda'_1,\lambda_2 \}$.
Here $\Omega_b h^2$ and $\Omega_c h^2$ are today's density of baryonic matter and cold dark matter respectively,
$100\theta_{\text{MC}}$ is 100 times the ratio of the angular diameter distance to the large scale structure sound horizon,
$\tau_{\text{re}}$ is the optical depth, $n_s$ is the scalar spectrum index, and $A_s$ is the amplitude of the power spectrum of primordial curvature perturbations.
We refer to CAMB and CosmoMC~\cite{Lewis:2002ah} and use the data combination of the latest CMB data released by the Planck collaboration in 2018, Planck 2018 TT,TE,EE$+$lowE$+$lensing \cite{Aghanim:2018eyx}, the BAO data including MGS~\cite{Ross:2014qpa}, 6DF~\cite{Beutler:2011hx} and DR12~\cite{Alam:2016hwk} and the PANTHEON sample consisting of 1048 SNIa measurements.
The results are summarized in the first column of Tab.I. 
$\lambda_2$ is $0.971_{-0.047}^{+0.043}$ and $\lambda'_1$ is $1.003\pm0.015 $ at 68$\%$ C.L..
According to Fig.~\ref{fig:LG}, it indicates that the Newton's constant $G=G_N$ is still acceptable both during the recombination epoch and in the late-time universe till now.
There is no evidence indicating the dynamic property of the Newton's constant.
The Hubble constant $H_0$ reads $67.78 \pm 0.48 $ km s$^{-1}$ Mpc$^{-1}$ at 68$\%$ C.L., which is in agreement with the result $67.4\pm 0.5 $ km s$^{-1}$ Mpc$^{-1}$ of Planck 2018.
The 68$\%$ limits for $M_B^1$ is $-19.363\pm 0.020$ mag, which is smaller than the previous constraint $-19.13\pm 0.01$~\cite{Zhao:2018gwk}.
The triangular plot of $\lambda'_1, M_B^1, \lambda_2, H_0, 100\theta_{\rm MC}$ and $n_s$ is also shown in Fig.\ref{fig:lambda12}.
$\lambda'_1$ has positive correlation with $M_B^1$ as shown in Eq.~(\ref{eq:mod}), but it's almost independent of other parameters.
By comparison, $\lambda_2$ is much more complicated.
It has strong and negative relationship with $100\theta_{\text{MC}}$  and $H_0$ due to its affect on $z_*$. 
The correlation between $\lambda_2$ and $n_s$ results from the transformation of $k \to k/\lambda_2$.
\begin{table}
\label{tb:result}
\caption{The 68$\%$ limits for the cosmological parameters in two models for different purpose. Notice that $\lambda_1(\lambda'_1)$ indicates $\lambda_1$ is a function of $\lambda'_1$.}
\begin{tabular}{p{3 cm}<{\centering}|p{4.5cm}<{\centering} p{4.5cm}<{\centering}   }
\hline
                & Probing the dynamics of $G$ with CMB, BAO and SNIa
                & Solving the $H_0$ tension with CMB, BAO and uncalibrated SNIa\\
\hline
$\Omega_b h^{2}$ & $0.02236\pm 0.00016$ & $0.02237\pm 0.00076$   \\

$\Omega_c h^{2}$ & $0.1197\pm 0.0010$   & $0.1189\pm0.0035$       \\

$100\theta_{\rm MC}$ & $1.04195_{-0.00144}^{+0.00143} $ &$1.04119_{-0.00206}^{+0.00208}$     \\

$\tau_{\rm re}$           & $0.055\pm 0.007$     & $0.056\pm0.007$         \\

$\ln(10^{10}A_s)$ &$3.043 \pm 0.015 $& $3.047_{-0.015}^{+0.014}$        \\
$n_s$           & $0.9629\pm 0.0064$     & $0.9665_{-0.0062}^{+0.0063}$         \\
\hline
$H_0$ [km s$^{-1}$ Mpc$^{-1}$]&$67.78 \pm 0.48$ &$67.62^{+1.24}_{-1.25}$  \\
\hline
$\lambda_0$              & $1$                  &$\ $                         \\
$\lambda_1$          &   $\lambda_1(\lambda'_1=1.003\pm0.015) $                 & $0.999\pm 0.014$                          \\
$\lambda_2$              & $0.971_{-0.047}^{+0.043}$                  &$\ $                         \\
\hline
   $M_B^1$          &   $-19.363\pm 0.020$              & -                    \\
\hline
\end{tabular}
\end{table}
   
\subsection{$H_0$ tension and Constant $G$ with Redshift}
\label{subsec:result-2}

Then, we try to solve the Hubble tension with a varying Newton's constant by consider an simple extension of $\Lambda$CDM model with another one free parameter $\lambda_0=\lambda_1=\lambda_2$. 
We use the data combination of Planck 2018 TT,TE,EE$+$lowE$+$lensing, the BAO data (6DF, MGS and DR12) and uncalibrated PANTHEON sample.
The results are shown in the second column of Tab.I: $\lambda_i = 0.999 \pm 0.014$ and $H_0=67.62^{+1.24}_{-1.25} \pm$ km s$^{-1}$ Mpc$^{-1}$ at 68$\%$ C.L.
Our results indicate that Newton's constant $G =(6.65635_{-0.18560}^{+0.18766} ) \times 10^{-11} \rm m^3kg^{-1}s^{-2}$ at 68$\%$ C.L., which is consistent with the value of CODATA 2018. 
However, the $H_0$ tension can't be solved with this method.

\section{Summary and discussion}
\label{sec:sum}

In this paper, we investigate how Newton's constant influences the CMB power spectra and the light curve of SINa. So the CMB data and SINa measurement can put a constraint on the Newton's constant.
Combining the Planck data released in 2018, the BAO data and PANTHEON samples, we run CAMB and CosmoMC with a varying $G$ during the recombination epoch and the redshift ranges of SINa measurement.
$G=G_N$ is located in the 68$\%$ C.L. ranges of the two periods.
We find no evidence of the dynamic property of Newton's constant.

In addition, considering the effect of Newton's constant on the expansion history of our universe, we have a try to solve the $H_0$ tension by freeing $G$ based on the $\Lambda$CDM model.
Adopting the combination of Planck 2018 TT,TE,EE+lowE+lensing+uncalibrated PANTHEON+BAO, we obtain $\lambda_i = 0.999 \pm 0.014$ and $H_0=67.62^{+1.24}_{-1.25}$ km s$^{-1}$ Mpc$^{-1}$ at 68$\%$ C.L..
With this method, the $H_0$ tension can't be solved.
At the same time, our results show that Newton's constant from this model is consistent with the value given by CODATA 2018.

\vspace{5mm}
\noindent {\bf Acknowledgments}
We acknowledge the use of HPC Cluster of Tianhe II in National Supercomputing Center in Guangzhou and HPC Cluster of ITP-CAS. We  would like to thank Qing-Guo Huang for his helpful discussions and advices on this paper.



\end{document}